\documentstyle[multicol,aps,psfig]{revtex}

\begin{document}     

%\draft

\preprint{aaaa}

\title{Complex singularities of the critical potential in the
large-$N$ limit}
\author{Y. Meurice{\footnote {e-mail:yannick-meurice@uiowa.edu}}  \\
{\it Department of Physics and Astronomy, The University of Iowa, 
Iowa City, Iowa 52242, USA{\footnote{permanent address}}}\\
and\\
{\it Fermilab, PO Box 500, Batavia,
Illinois 60510-0500, USA}}

\maketitle
\begin{abstract}
We show with two numerical examples that the conventional expansion 
in powers of the field for  
the critical potential
of 3-dimensional $O(N)$ models in the 
large-$N$ limit, does not converge for values of $\phi ^2$
larger than some  critical value. This can be explained by the existence of
conjugated branch points in the complex $\phi ^2$ plane.
Pad\'e approximants $[L+3/L]$ for the critical potential 
apparently converge at large
$\phi^2$. This allows high-precision
calculation of the fixed point in a more 
suitable set of coordinates. 
We argue that the singularities are
generic and not an artifact of the large-$N$ limit. We show that
ignoring these singularities 
may lead to inaccurate approximations.

\end{abstract}

%\pacs{PACS: 11.10.-z, 11.15.Bt, 12.38.Cy, 31.15.Md}
\begin{multicols}{2}\global\columnwidth20.5pc
%\multicolsep=8pt plus 4pt minus 3pt
 
\section{Introduction}
Since the early days of the 
renormalization group (RG) method\cite{wilson}, 
3-dimensional scalar models 
have been identified 
as an important laboratory to discuss the
existence of non-trivial fixed points and 
the large cut-off (or small lattice spacing)
limit of field theory models.
In the case of $N$-components models with 
an $O(N)$ invariant Lagrangian, the RG transformation
becomes particularly simple in the large-$N$ limit \cite{ma73}.
The construction of the effective potential for these models 
is discussed in Refs. \cite{col,town,app}.
Later, motivated by perturbative results indicating
the existence of an UV stable tricritical 
fixed point for $N$ large enough\cite{pisarski}, a new mechanism 
allowing spontaneous breakdown of scale invariance and dynamical 
mass generation was found in the large-$N$ limit\cite{bmb}.
In the following, we call this mechanism the ``BMB mechanism''.
It was argued\cite{amit84} that the BMB mechanism is compatible with
a zero vacuum energy and a better understanding 
of this question might suggest
a solution to the cosmological constant problem.
Spontaneous breaking of scale invariance 
is also discussed \cite{bardeen} with related methods in four-dimensional
models of clear interest in the context of particle physics.
However, doubts were cast\cite{david84} about the fact that the 
BMB mechanism is generic
and it is commonly believed that it disappears at finite $N$.

In this article we report results which force us to reconsider 
the way we think about non-trivial fixed points.
We usually think of the RG flows as taking place in a
space of bare couplings or more generally in a space of functions.
The necessity for this 
more general point of view appears quite clearly in exact renormalization
group equations \cite{exact}. Unfortunately, it seems impossible 
to decide {\it a-priori} 
which space of functions should be considered 
to study the RG flows.
It is clear from perturbation theory that near the Gaussian
fixed point, low dimensional polynomial approximations 
of the local potential should be adequate.
However, it is not clear that this kind of 
approximation should be valid
far away from the Gaussian fixed point and in particular 
near non-trivial fixed points. 

In the following, we concentrate on the non-trivial fixed point 
found numerically in the case $N=1$ by K. Wilson \cite{wilson}. 
This fixed point is located on a hypersurface 
of second order phase transition which separates the symmetric phase from 
the broken symmetry phase. 
In the following we call this fixed point the Heisenberg fixed 
point (HFP for short) as in Ref. \cite{david84}. 
It should not be confused with the fixed point relevant for the 
BMB mechanism and which is not studied in detail here.
The main result of the article is 
that the bare potential corresponding to the HFP
has singularities in the 
complex plane and that
ignoring these singularities may 
lead to inaccurate approximations. 
These claims are based on explicit calculations 
performed in the large-$N$ limit for two $O(N)$ invariant models
reviewed in section
\ref{sec:models}. These two models are: 
1) a model with a $k^2$ kinetic term
together with a sharp cut-off, the sharp cut-off model (SCM) for
short; 2) Dyson's hierarchical model (HM)\cite{dyson,baker}. 
%In both cases,
%it is possible to write 
%the inverse of the first derivative of the bare
%potential corresponding to the non-trivial fixed point in close form. 

Before entering into
technical details, three points should be clear. First, all the results 
presented here are based on the analysis of long numerical series
and no attempt is made to give rigorous proofs. Second, 
in order to understand some of the statements made below, 
the reader should be
aware that even though, at leading order in the large-$N$ approximation, 
the critical exponents take $N$-independent 
values, the same
approximation provides finite $N$ approximate HFP which are
$N$-dependent.
A more precise formulation of this statement can be found in
Sections \ref{sec:models} and  \ref{sec:fp5}.
Third, we only work in 3 dimensions. The precise meaning of this
statement for the hierarchical model is explained at the end of
section \ref{sec:models}.

In section \ref{sec:hfp}, we review the basic equations 
\cite{ma73,david84} defining the 
HFP for the SCM. We then show that the definition can be extended 
naturally for the HM. The correctness of this definition is
verified later in the paper.
In section \ref{sec:series}, we present the methods used to calculate
the critical potential expanded as a Taylor series in $\phi^2$.
The coefficients of this expansion are called the critical couplings.
The main conclusion that we can infer 
from our numerical results 
is that the Taylor series
is inadequate for large values of $\phi^2$. First of all, one
half of the critical couplings are negative. If we truncate the Taylor
series at an order such that the coefficient of the highest order is 
negative, we obtain an ill-defined functional integral.
In addition, the absolute value of the critical couplings grows
exponentially with the order and the expansion has a finite radius of
convergence. Consequently, the idea that the
critical potential associated with the HFP 
can be approximated by polynomials should be
reconsidered.

It is nevertheless possible to define the critical theory by using 
Pad\'e approximants for the critical potential. In section
\ref{sec:pade}, we show that sequences of
approximants converge toward the expected function 
in a way very reminiscent of the
case of the anharmonic oscillator were the convergence 
can be proven rigorously \cite{loeffel69}.
In addition, the zeroes and the poles of these
approximants are located far away from the real positive axis and 
follow patterns that strongly suggest the existence of two complex
conjugated branch points. 

The complex singularities of the critical potential should not be
interpreted as a failure of the RG approach but rather as an 
artifact of the system of coordinates used. In section \ref{sec:fp5},
we present consistent arguments showing 
that in a different system of coordinates \cite{kw,guide}, the 
function associated with the HFP is free of singularities.
In this system of coordinates, finite dimensional truncation 
is a meaningful procedure which, in the case of the HM,  
allows comparison 
with independent numerical calculations at finite $N$.  
An example of such a calculation is presented in the case $N=5$.

In section \ref{sec:disc}, 
we discuss the errors associated with two 
approximate procedures that can be used to deal with the singularities.
The first procedure which is justified in the context of perturbation 
theory and does not require an understanding of the 
singularities, consists in 
truncating the potential at order $(\phi^2)^3$. The second 
procedure consists 
in restricting the range of integration of $\phi^2$ to the radius 
of convergence of the critical potential. 
If the range of integration is large enough, this second procedure
generates small errors \cite{convpert}.
As far as the calculation of the HFP in the system of coordinates 
of Section \ref{sec:fp5} is concerned, 
both procedures have a low accuracy for both models 
considered.
In the conclusions, we explain why we believe that the singularities
persist at finite $N$ and we discuss implications 
of the existence of these singularities for other problems.

\section{Models}
\label{sec:models}

We consider lattice models defined by the partition function
\begin{equation}
Z(\vec{J})=\prod _x\int d^N\phi_x {\rm e}^{-S
+\sum_x\vec{J}_x\vec{\phi}_x}\ ,
\end{equation}
with
\begin{equation}
S=-{1\over 2}\sum_{xy}
\vec{\phi}_x\Delta_{xy}\vec{\phi}_y+\sum_xV_o(\phi^2_x)\ .
\end{equation}
We use the notation $\phi^2_x \equiv \vec{\phi}_x.\vec{\phi}_x$ and 
$\Delta_{xy}$ is a symmetric
matrix with negative eigenvalues, such as discrete versions of the 
Laplacian. For the simplicity of the
presentation, we will assume that $\sum_x\Delta_{xy}=0$. If it is not the
case, one can always subtract the zero mode from $\Delta$ and compensate it
with a new term in $V$.

Defining the rescaled potential
\begin{equation}
V_0(X)=NU_0({X\over N})\ ,
\label{eq:u0}
\end{equation}
and performing a Legendre transform from the source $\vec{J}$ to the 
external classical field $\vec{\phi}_c$, one can show that 
in the large $N$ limit\cite{david84} that
$M^2\equiv 2\partial V_{eff}/\partial \phi^2_c$ obeys the self-consistent
equation 
\begin{equation}
2U_0'(\phi_c^2+f_{\Delta}(M^2))=M^2\ ,
\end{equation}
where $f_{\Delta}(M^2)$ is the one-loop integral corresponding to the
quadratic form $\Delta$ and a mass term $M^2$. 
The prime denotes the derivative with respect to the $O(N)$ invariant 
argument.
The explicit form of
$f$ for the two models discussed in the following are given in Eqs. 
(\ref{eq:fscm}-\ref{eq:fhm}). 
Precise definitions of $\phi ^2_c$ and the effective potential 
$V_{eff}$ are given in \cite{david84}.
 
Up to now, all the quantities introduced are dimensionless.
They can be interpreted as dimensionful quantities expressed in 
cut-off units.
Let us consider two models, the first one with a rescaled potential
$U_0$, a UV cutoff $\Lambda$  and a quadratic form $\Delta$ 
and a second  model with a rescaled potential
$U_{0,S}$, a UV cutoff $\Lambda/S$ and a quadratic form $\Delta_S$. 
For $D=3$ and in the large-$N$ limit, the two models have the same 
dimensionful zero-momentum Green's functions provided that:
\begin{eqnarray}
\label{eq:rg}
&U&'_{0,S}(\phi^2)= \\ 
&S&^2U_0' \biggl(\bigl(\phi^2-f_{\Delta_S}(2U'_{0,S}(\phi^2)\bigr)/S+
f_{\Delta}\bigl((2/S^2) U'_{0,S}(\phi^2)\bigr)\biggr)\nonumber
\end{eqnarray}
In two special cases, the dimensionless expression for the one-loop
diagram is independent of the cut-off. In other words,
$f_{\Delta}=f_{\Delta_S}\equiv f$ and the fixed point equation becomes
very simple \cite{ma73,david84}.

We now discuss the two models where this simplification occurs.
In the SCM, $\Delta$ becomes $k^2$ in the momentum
representation (Fourier modes). The momentum cutoff is sharp: $k^2\leq 1$ (in 
cut-off units). This is why we call this model 
the sharp cutoff model.
The non-renormalization of the kinetic term is justified in the 
large-$N$ limit \cite{david84}. For this model,
\begin{equation}
f_{SCM}(z)=\int _{|k|\leq 1}{{d^3 k}\over {(2\pi)^3}}{1\over {k^2+z}}\ .
\label{eq:fscm}
\end{equation}

By construction \cite{baker}, 
the kinetic term of Dyson's hierarchical model (HM)
is not renormalized and we have  
\begin{equation}
f_{HM}(z)=\sum_{n=0}^{\infty}{{2^{-n-1}}\over{b(c/2)^n+z}}\ ,
\label{eq:fhm}
\end{equation}
with $c=2^{1-2/D}$ and $b$=${\beta c\over{2-c}}$. 
The inverse temperature $\beta$ and the parameter 
$c$ appear in the hamiltonian of the HM in a way that 
is explained in section II of Ref. \cite{gam3}. The parameter $c$ 
is related to the dimension $D$ by considering the scaling of a 
massless Gaussian field. In the following we will consider the 
case $c=2^{1/3}$ ($D=3$) exclusively. 
In addition, $\beta$ will be set to 1 as in
other fixed point calculations \cite{gam3}.
Different values of $\beta$ can be introduced by a trivial rescaling.
Note also that the cutoff cannot be changed continuously for the
HM, because the invariance of $f$ is only valid when we integrate 
the degrees of freedom of a the largest momentum shell (corresponding 
to the hierarchically nested blocks in configuration space)
all at the same time.
For the HM, the density of sites is reduced by a factor 2 at each RG 
transformation. The linear dimension (``lattice spacing'') 
is thus increased by by a factor $2^{1/D}$ and the cutoff decreased by
the same factor.
Consequently, for the HM, 
Eq. (\ref{eq:rg}) should be understood only with $S=2^{q/D}=2^{q/3}$ 
and $q$ integer. 

\section{The HFP}
\label{sec:hfp}

In this section, we review the construction of the HFP for the SCM,
and we show that the construction can be extended in a natural (but
non-trivial) way for the HM. The first step consists in finding all
the fixed points of the RG Eq. (\ref{eq:rg}).
Following Refs. \cite{ma73,david84}, we introduce the inverse function: 
\begin{equation}
F(2U'_0(\phi^2))=\phi^2\ ,
\end{equation}
and the function $H(z)\equiv F(z)-f(z)$ ,
%\begin{equation}
%\end{equation}
where the one-loop function $f$ has been defined in the previous section for
the two models considered. With these notations, 
the fixed point equation corresponding to Eq. (\ref{eq:rg}) is simply 
\begin{equation}
H(z)=SH(z/S^2)\ .
\label{eq:fpeqh}
\end{equation}

For the SCM, $S$ is allowed to vary continuously in Eq. (\ref{eq:fpeqh})
and the general solution is 
\begin{equation}
F(z)=f_{SCM}(z)+Kz^{1/2}\ .
\end{equation}
For the HM, $S$ can only be an integer power of $2^{1/3}$ and the
general solution has an infinite number of free parameters:
\begin{equation}
F(z)=f_{HM}(z)+\sum_qK_q z^{1/2+iq\omega}\ ,
\label{eq:hmfp}
\end{equation}
with
\begin{equation}
\omega\equiv {3\pi\over{{\rm ln}2}}\simeq 13.6\ ,
\label{eq:omega}
\end{equation}
and $q$ runs over positive and negative integers.
The only restriction on the constants $K$ and $K_q$ is that $F$ should
have a well defined inverse which is real when $F (=\phi^2)$ is real and
positive.

It is clear from Eqs. (\ref{eq:fscm}) and (\ref{eq:fhm}) that for both
models $f(z)$ has singularities along the negative real axis and that, in
general, $F(z)$ cannot be defined for $z$ real and negative. 
This imposes restrictions on the choice of the constants $K$ and $K_q$.
For instance, in the case of 
the SCM, when $K$ takes a large positive value, it 
is impossible to reach small values of $F=\phi^2$ when $z\geq 0$ and
the fixed point has no obvious physical interpretation. However, there
is a special positive value of $K$ for which the singularity of
$f_{SCM}$ is exactly canceled and an analytic continuation for $z<0$
is possible. Its exact value can be calculated \cite{david84} 
by decomposing $f_{SCM}$ into 
a regular part $f_{SCM,reg.}$ and a singular part $f_{SCM,sing.}$. 
Using elementary trigonometric identities, one finds 
\begin{equation}
f_{SCM,reg.}(z)={1\over{2\pi^2}}(1+z^{1/2}{\rm Arctan}(z^{1/2}))\ ,
\end{equation}
and 
\begin{equation}
f_{SCM,sing.}(z)=-{1\over{4\pi} }z^{1/2}\ .
\end{equation}
Consequently, if we choose $K={1\over{4\pi}}$, $F$ reduces to $f_{SCM,reg.}$.

A more detailed analysis \cite{david84}, shows that this value of $K$
is the only positive value of $K$ for which $U'$ can be defined for 
any real positive $\phi^2$. On the other hand, for negative $K$, one 
obtains a line of fixed points ending (for $K=0$) at the fixed point
relevant for the BMB mechanism.
Given the isolation of the fixed point with
$K={1\over{4\pi}}$, it is easy to identify it with the HFP. We denote the 
corresponding inverse function by $F^{\star}_{SCM}(z)=f_{SCM,reg.}(z)$.
As promised this function is analytical in a neighborhood of the
origin and has a Taylor  expansion:
\begin{equation}
F^{\star}_{SCM}(z)={1\over{2\pi^2}}(1+z-{z^2\over 3}+{z^3\over
  5}+\dots)
\end{equation}
This expansion has a radius of convergence equal to 1 due to
a logarithmic singularity at $z=-1$. However, as we will see in
section \ref{sec:series}, this expansion allows us to construct an 
inverse power
series and $U_0$.

In the case of the HM, the decomposition into a regular and singular 
part is more tedious. Fortunately, this problem is a particular case 
of a problem solved in section 5 of Ref. \cite{osc} where
Eq. (5.6) with $A=c^2$, $B=c^{-1}$ and $f(z)=G(z/b)$ yields
\begin{equation}
f_{HM,sing.}(z)=-{\omega \over{4b}}\sum_q{\bigl( {z/b}\bigr)
  ^{1/2+iq\omega}
\over
  {\rm sin}(\pi(1/2+iq\omega))}\ ,
\label{eq:fhms}
\end{equation}
with $b$ and $c$ defined in section \ref{sec:models}.

If we compare this expression with the general solution of the fixed
point equation (\ref{eq:hmfp}),
we see that in both expressions, the power  
$z^{1/2+iq\omega}$ appears for all positive and negative 
integer values of $q$. There exists a unique choice of the $K_q$ in Eq. 
(\ref{eq:hmfp}) which cancels exactly the singular part of $f_{HM}$.
We call the corresponding fixed point the HFP of the HM.
The numerical closeness with the finite $N$ HFP discussed in section
\ref{sec:fp5} 
confirms the validity of this analogic definition.
We call the corresponding inverse function $F_{HM}^{\star}$. Using
Eq. (5.5) of Ref. \cite{osc}, we find 
\begin{equation}
F_{HM}^{\star}(z)=f_{HM,reg.}={1\over{2b}}\sum_{l=0}^{\infty}\biggl({-z\over
  b}\biggr)^l{1\over{1-c^{2l-1}}}\ .
\label{eq:fexp}
\end{equation}
This expansion has a radius of convergence $bc^2=2.7024\dots$ for the
choice
of parameters used here.

It is possible to check the accuracy of the expansion given in Eq. 
(\ref{eq:fexp}) by using the identity 
$F_{HM}^{\star}(z)=f_{HM}(z)-f_{HM,sing.}$.
Note that the two terms of the r.h.s. cannot be defined separately on
the negative real axis. On the real positive axis, $f_{HM,sing.}$ is
dominated by the $q=0$ term. Numerically, 
\begin{equation}
K_0={3\pi\over{4b^{3/2}{\rm ln}2}}=1.530339\dots .
\label{eq:ko}
\end{equation}
The terms with $q=\pm 1$ produce log-periodic oscillations of 
amplitude $1.7 \times 10^{-18}$. The terms with larger $|q|$ have a
much smaller amplitude. These findings are consistent with the log-periodic
oscillations found numerically in HT expansions \cite{osc1,osc}.
The oscillatory terms are very small along the
positive real axis. However, in the complex plane, if we write $z=r{\rm
  e}^{i\theta}$, the amplitude is multiplied by ${\rm
  e}^{-q\omega\theta}$
which compensates the suppression of the denominators in Eq. (\ref{eq:fhms}), 
if $\theta \rightarrow +\pi (-\pi)$
when $q<0$ ($q>0$). In conclusion, along the real positive axis, we can 
use the approximation
\begin{equation}
F_{HM}^{\star}(z)\simeq f_{HM}(z)+K_0 z^{1/2}\ ,
\label{eq:fstarapp}
\end{equation}
%with $K_0$ given in Eq. (\ref{eq:ko}).
with an accuracy of 18 significant digits, but this approximation is 
certainly not valid near the negative real axis.

\section{Calculation of the critical potential $U_0^{\star}$}
\label{sec:series}

\begin{figure}
\centerline{\psfig{figure=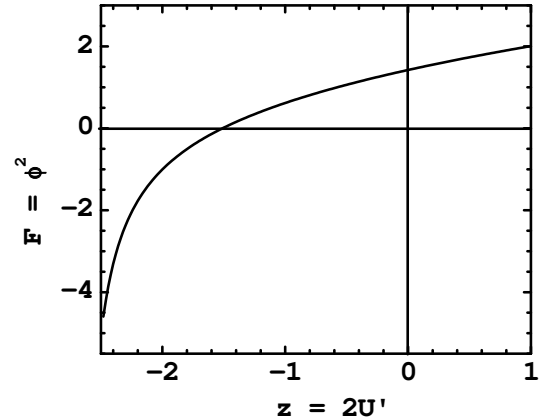,width=3.3in}}
\vskip10pt
\caption{$F_{HM}^{\star}(z)$ versus $z$ }
\label{fig:foz}
\end{figure}

In the previous section, we have provided power series for 
the inverse
function $F(z)$ corresponding to the HFP of the SCM and the HM. 
We can use these series to define $F(z)$ on the negative real axis. 
In both cases, as we move toward more negative values of $z$, $F$ becomes 
zero within the radius of convergence of the expansion. The situation
is illustrated in Fig. \ref{fig:foz} for the HM. 
Numerically, we have $F_{HM}^{\star}(-1.5107\dots)=0$ and 
$F_{SCM}^{\star}(-0.6948\dots)=0$. We then reexpand the series
about that value of $z$ (which corresponds to $F=\phi^2=0$) and invert
it. The resulting series is an expansion of $2U_0^{\star}$' in $\phi^2$.
After integration, and up to an arbitrary constant $u_0$, we obtain a 
Taylor series for the critical potential $U_0^{\star}$ corresponding 
to the HFP. We denote the expansion as 
\begin{equation}
U_0^{\star}(\phi^2)=\sum_{n=0}^{\infty} u_n (\phi^2)^n\ .
\label{eq:uudef}
\end{equation}
The precise determination of the zero of $F$ is obtained by Newton's
method with a large order polynomial expansion. This expansion is then
reexpanded about the zero and the large order coefficients in the original
expansion have an effect on the low order coefficients 
of the reexpanded series. 
We have checked that the order was sufficiently large 
to stabilize the results presented hereafter. 
\begin{figure}
\centerline{\psfig{figure=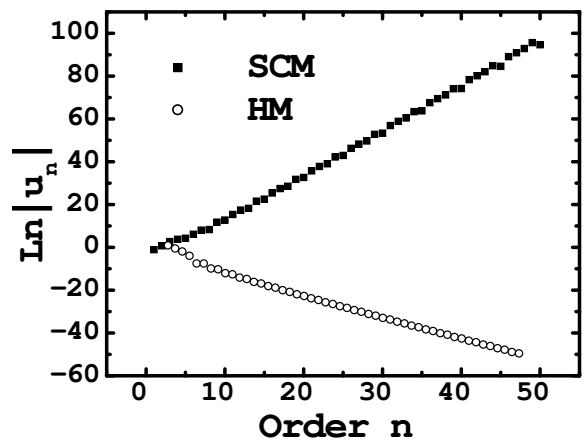,width=3.3in}}
\vskip10pt
\caption{Natural Logarithm of the absolute value of the coefficients
$u_n$ of the critical potential $U_0^{\star}$ defined in Eq. (\ref{eq:uudef}) 
for the SCM 
(filled squares) and the HM (empty circles).}
\label{fig:lnco}
\end{figure}

The absolute values of the first 50 coefficients of both models 
are shown in Fig. \ref{fig:lnco}.
In both cases, it appears clearly that the absolute value grows at an
exponential rate. Linear fits of the right part of Fig. \ref{fig:lnco}
suggest a radius of convergence of order 0.11 for the SCM and 2.5 for
the HM. The signs of both series follow the periodic pattern: + - - + +
- + + - -
for the SCM and + + - - for the HM. This suggests singularities in the
complex plane at an angle ${k\pi\over 5}$ with respect to the positive
real axis ($k=$ 1, 3, 7, 9) for the SCM and along the imaginary axis
for the HM. This analysis is confirmed by an analysis of the poles 
of Pad\'e approximants presented in the next section.

\section{Pad\'e approximants of $U_0^{\star}$}
\label{sec:pade}

\begin{figure}
\centerline{\psfig{figure=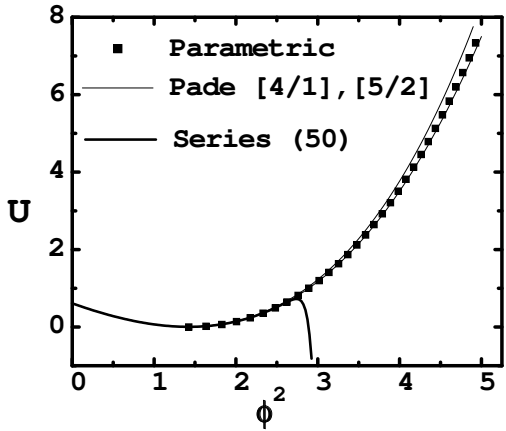,width=3.3in}}
\vskip10pt
\caption{$U_0^{\star}(\phi ^2)$ for the HM 
with a parametric plot (filled squares), the
series truncated at order 50 (thick solid line) and Pad\'e
approximants [4/1] (thin line slightly above the squares) and 
 [5/2] (thin line closer to the squares). The constant has been fixed
 in such a way that the value at the minimum is zero.}
\label{fig:uuu}
\end{figure}

At this point, our series expansion of the critical potential does 
not allows us to define the critical theory as a functional integral.
As $\phi^2$ exceeds the critical values estimated in the previous
section, the power series is unable to reproduce the expected function
$U_0^{\star}$. The situation is illustrated in Fig. \ref{fig:uuu} for the HM.

The numerical values of $U_0^{\star}$ in Fig. \ref{fig:uuu} 
have been calculated
using a parametric representation (with $z$ as the parameter). 
We have calculated pairs of 
values
\begin{equation}
\left(F^{\star}(z),{1\over 2}\bigl(zF^{\star}(z)-\int_0^z
dz'F^{\star}(z')\bigr)\right)\ , 
\label{eq:para}
\end{equation}
for various real positive values of $z$.  
A simple graphical analysis performed by representing $U_0^{\star}$ as a surface
on Fig. \ref{fig:foz} shows that each pair of values in 
in Eq. (\ref{eq:para}) corresponds to 
a pair $(\phi^2, U_0^{\star}(\phi^2))$ with the arbitrary constant
in $U_0^{\star}$ fixed in such a way that $U_0^{\star}$ vanishes at its minimum. 
We have calculated 
$F^{\star}$ by using the independent but approximate Eq. (\ref{eq:fstarapp}).
As explained in section \ref{sec:hfp},
the approximate expression is only valid 
for $z$ real and positive and should give 18 correct significant digits. 
In Fig. \ref{fig:uuu}, we have used the values $z=2U'_0=0,\ 0.25, 0.5,
\dots$.
This is why the filled 
squares only appear when the derivative of $U_0^{\star}$ is positive.
Unlike Eq. (\ref{eq:fexp}) which has a radius of convergence 1, the
approximate expression  Eq. (\ref{eq:fstarapp}) remains valid 
for large positive values of $z$. It is thus
possible to check if Pad\'e approximants can be used to represent 
the critical potential beyond the radius of convergence of its 
Taylor expansion. Fig. \ref{fig:uuu} shows that low order approximants
are close to the parametric curve. As the order increases, the curves
coalesce with the parametric curve and a more refined description is
necessary. 

\begin{figure}
\centerline{\psfig{figure=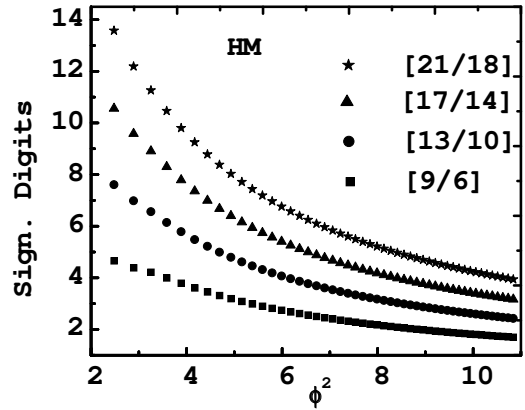,width=3.3in}}
\vskip10pt
\caption{Number of correct significant digits obtained with Pad\'e 
approximants [9/6], [13/10], [17/14] and [21/18] 
for various values of $\phi ^2$ for the HM.}
\label{fig:padeconv}
\end{figure}

In Fig. \ref{fig:padeconv}, we give the accuracy reached 
by various approximants for the HM with a broad range of $\phi^2$ 
(more than 4 times the radius of convergence). As the order of
the approximants increase the accuracy increases but at a rate which 
is slower for larger values of $\phi^2$. The figure is very similar 
to sequences of Pad\'e approximants obtained for the 
ground state of the anharmonic oscillator
(see Fig. 1 of Ref. \cite{convpert}),
where the convergence can be proven rigorously \cite{loeffel69}.
Note that the slow convergence at large $\phi^2$ 
is not a serious problem, since the 
contributions for 
large $\phi^2$ are exponentially suppressed in the functional integral.
The choice of $[L+3/L]$ approximants is discussed in more detail
below.
Up to now, we only discussed the HM. Following the same procedure for 
the SCM, we obtain very similar figures (with a different $\phi^2$
scale)
which we have not displayed.

\begin{figure}
\centerline{\psfig{figure=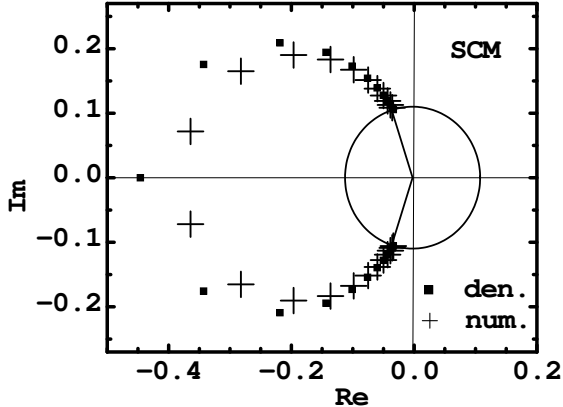,width=3.3in}}
\vskip10pt
\caption{Real and imaginary parts of the roots of the denominator 
(filled squares) and numerator (crosses) of 
a [26/23] Pad\'e approximant for the SCM  
The solid circle has a radius 0.11 and the
two solid lines make angles $\pm {3\pi\over 5}$ with respect to the
positive real axis.}
\label{fig:rootscm}
\end{figure}

\begin{figure}
\centerline{\psfig{figure=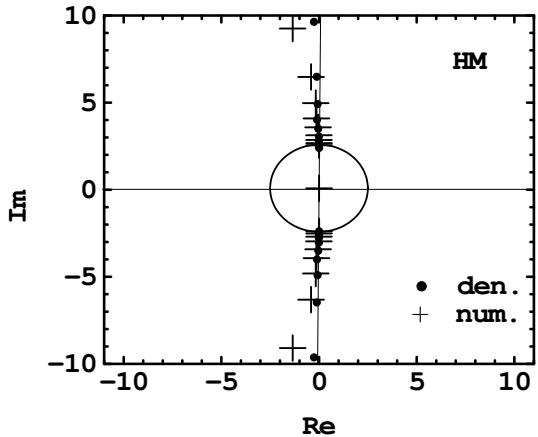,width=3.3in}}
\vskip10pt
\caption{Real and imaginary parts of the roots of the denominator 
(filled circles) and numerator (crosses) of 
a [26/23] Pad\'e approximant for the HM. 
The solid circle has a radius 2.5. Two roots farther away on the
imaginary axis and one root farther away near the negative real axis 
are not displayed.}
\label{fig:roothm}
\end{figure}

The singularities of $U_0^{\star}$ in the complex $\phi^2$ plane can be
inferred from the location of the zeroes and poles of the Pad\'e 
approximants. As $L$ becomes large, regular patterns appear.
Examples are shown in Fig. \ref{fig:rootscm} for the SCM and 
Fig. \ref{fig:roothm} for the HM.
In both cases, the zeroes and 
poles approximately alternate along two lines 
ending where singularities were expected from the analysis of
coefficients
in Section \ref{sec:series}. This pattern suggests \cite{baker96} 
the existence of two
complex conjugated branch points at the end of these lines.


\begin{figure}
\centerline{\psfig{figure=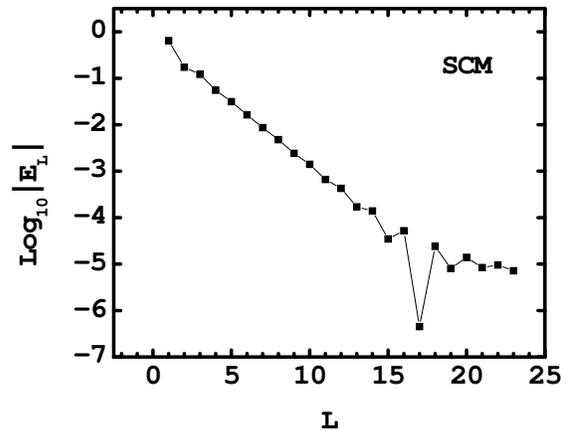,width=3.3in}}
\vskip10pt
\caption{Log$|E_L|$ versus $L$.}
\label{fig:sdco}
\end{figure}

The choice of $[L+3/L]$ approximants is easily justified for the 
SCM. At large $|z|$, $f_{SCM}(z)\propto 1/z$ and and
$F^{\star}_{SCM}(z)
\simeq {1\over{4\pi} }z^{1/2}$. For large $|\phi^2|$, 
$U_0^{\star}$'$\simeq 8\pi^2 (\phi^2)^2$ and 
$U_0^{\star}\simeq{8\pi^2\over{3}} (\phi^2)^3$. 
Consequently a $[L+3/L]$ approximant should have the correct asymptotic
behavior. More precisely, 
if $a_{L+3}$ and $b_L$ are the leading coefficients of the numerator
and
denominator of a Pad\'e $[L+3/L]$ respectively, we expect that when $L$ is
large
\begin{equation}
{a_{L+3}\over b_L}\rightarrow{8\pi^2\over 3}
\label{eq:limas}
\end{equation}
Defining a quantity 
\begin{equation}
E_L\equiv 1-{{3a_{L+3}}\over{8\pi^2 b_L}}\ ,
\end{equation}
that measures the departure from the expected asymptotic behavior, we
see from Fig. \ref{fig:sdco} that as $L$ increases, the
discrepancy
diminishes exponentially.

In the case of the HM,  the situation is more 
intricate. From Eq. (\ref{eq:fstarapp}), we may be tempted to
conclude that the two cases are similar. Unfortunately, Eq. (\ref{eq:fstarapp})
is a real equation not a complex one. In the complex plane, the terms 
with $q\neq 0$ become important near the negative real axis and no
simple simple limit as in Eq. (\ref{eq:limas}) applies. However,
if  we need $U_0^{\star}$ only along the real positive axis, Fig. \ref{fig:padeconv}
justifies the use of the $[L+3/L]$ sequence of approximants.

\section{The HFP in a convenient set of coordinates}
\label{sec:fp5}

As explained in the introduction
we can think that the RG flows move in a space of functions.
The system of coordinates for this space can be chosen in a way which
is convenient to make approximations. A particularly convenient system
of coordinates for the HM consists in considering the 
Fourier transform of the local measure of integration \cite{kw,guide}.
In this system of coordinates and at leading order in
the $1/N$ expansion, the HFP for a given $N$ reads: 
\begin{equation}
R^{\star}(\vec{k})\propto \int d^N\phi {\rm e}^{-{b\over 2}\phi^2-
NU_0^{\star}(\phi^2/N)+i\vec{k}.\vec{\phi}}\ .
\label{eq:fpint}
\end{equation}
The quadratic term proportional to $b$ is due to the fact that 
the quadratic form $\Delta$ for the HM has a zero mode.
We then Taylor expand
\begin{equation}
R^{\star}(\vec{k})=1+\sum_{n=1}^{\infty}a_n (k^2)^n\ ,
\label{eq:fp}
\end{equation}
and consider the $a_n$ as the our new set of coordinates.
The advantage of this representation is that it is possible 
to make very accurate calculations by using polynomial 
approximations \cite{kw,guide,gam3} 
of the infinite sum in Eq. (\ref{eq:fp}). In this section and the 
next section, 
we discuss the details of the calculations for the HM.
The case of the SCM shares many similarities with the HM 
and is discussed briefly at the 
end of each section.

\begin{figure}
\centerline{\psfig{figure=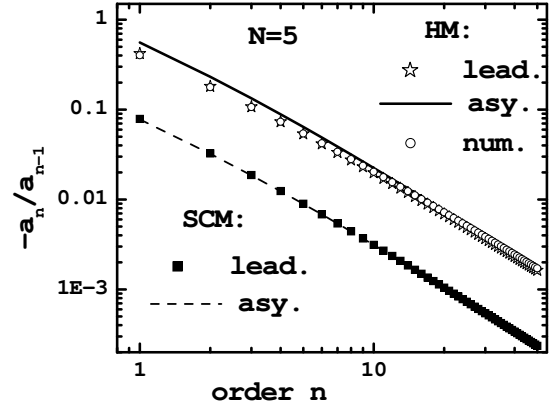,width=3.3in}}
\vskip10pt
\caption{Ratios of successive coefficients for the HM, using 
the leading order Eq. (\ref{eq:fpint}) (stars), the asymptotic 
formula Eq. (\ref{eq:asympratios}) (continuous line) and the 
numerical fixed point (empty circles). Same results for the SCM: leading
order (filled squares) and asymptotic (dashed line). In all cases, $N=5$.}
\label{fig:decayrat}
\end{figure}

We have performed a numerical calculation of the $a_n$ of the HM
using Eq. (\ref{eq:fpint}) in the 
particular case $N=5$. The study of the ratios of successive
coefficients displayed in Fig. \ref{fig:decayrat} indicates that 
the $|a_n|$ decay faster than $1/n!$ and that $R^{\star}(\vec{k})$ is 
analytical over the entire complex $k^2$ plane in contrast to
$U_0^{\star}(\phi^2)$ which has a finite radius of convergence in the complex
$\phi^2$ plane.

The good convergence of $R^{\star}(\vec{k})$ can be explained 
by an approximate calculation. 
The $\phi$ integral that is performed in the calculation
of the $a_n$ has a positive integrand with a peak moving to larger 
values of $|\phi|$ when $n$ increases. For sufficiently large values
of $n$, we can replace $U_0^{\star}$ by its asymptotic behavior on the 
positive real $\phi^2$ axis 
which can be derived from the approximate Eq. (\ref{eq:fstarapp}) for the HM:
\begin{equation}
R^{\star}(\vec{k})\sim \int d^N\phi {\rm e}^{-(1/(6N^2K_0^2))
(\phi^2)^3+i\vec{k}.\vec{\phi}}\ .
\end{equation}
With this approximation, the $a_n$ can be expressed  exactly in terms of 
gamma functions and a simple calculation yields
\begin{equation}
-{a_n\over{a_{n-1}}}\simeq {{(6N^2K_0^2)^{1/3}\Gamma((N+2n)/6)}\over{4n(n-1+N/2)
\Gamma((N+2(n-1))/6)}}\ .
\label{eq:asympratios}
\end{equation}
Note that there are no free
parameters
in this formula.
Fig. \ref{fig:decayrat} shows that Eq. (\ref{eq:asympratios}) is a
very good approximation of the ratios obtained numerically from
Eq. (\ref{eq:fpint}). 

We have also calculated the $a_n$ corresponding to the HFP for $N=5$ 
using the numerical method developed in the case $N=1$ 
in Ref. \cite{gam3} and which
can be extended easily for arbitrary $N$. In brief, 
it consists of finding the stable manifold by fine tuning the
temperature and then iterating the RG transformation 
in order to get rid of the irrelevant directions. This procedure is 
very accurate and completely 
independent of the approximations made in this article. Remarkably, 
we found that even though $N=5$ is not a large number, the first
coefficients obtained in the leading order in the $1/N$ approximation
coincide with about two significant digits with the accurate values 
found numerically with $N=5$. 
As the order increases, the accuracy
degrades slowly. This is explained in more detail below. However, 
the ratios of successive coefficients still follows 
closely the asymptotic
prediction obtained from Eq. (\ref{eq:asympratios}). 
This strongly suggests that the $(\phi^2)^3$ asymptotic behavior of the
critical potential persists at finite $N$.

Except for the comparison with independent numerical calculations at finite 
$N$, the same calculations can be performed for the SCM with minor changes 
($b\rightarrow 0$ and $K_0\rightarrow K$). The results are also shown in 
Fig. \ref{fig:decayrat} where one can see that the agreement with the
asymptotic formula is very good even at low order.

\section{Discussion of alternate procedures}
\label{sec:disc}
\begin{figure}
\centerline{\psfig{figure=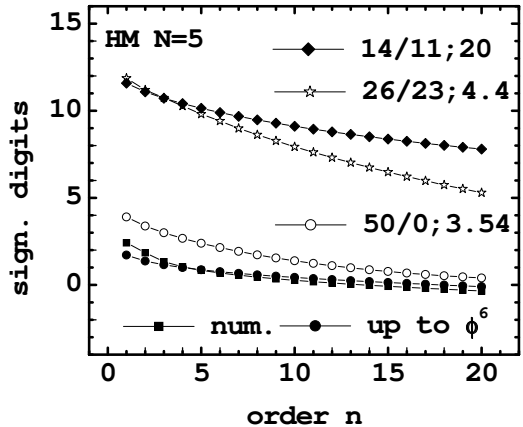,width=3.3in}}
\vskip10pt
\caption{Number of significant digits common with our best estimate
for the $a_n$ obtained for the HM 
from Eq. (\ref{eq:fpint}) with $n$ (the order) 
going from 1 to 20.  The alternate procedure are the truncation at 
order $(\phi^2)^3$ (filled circles), the $N=5$ accurate 
numerical result (filled
square), no Pad\'e approximants but a truncation of the range of
integration close to the radius of convergence (empty circles), 
a restriction of the range of integration for $\phi<4.4$ (stars), and
a [14/11] Pad\'e (diamonds).}
\label{fig:various}
\end{figure}

In section \ref{sec:pade}, we have shown that the Pad\'e approximants
provide accurate values of $U_0^{\star}$ far beyond its radius of
convergence.
In order to estimate the error 
on the new coordinates $a_n$ 
due to the 
use of approximants for 
$U_0^{\star}$, we can vary the range
of integration and change the approximants. For instance the 
values of $a_n$ of the HM 
used in Fig. \ref{fig:decayrat} have been calculated
using a range of integration $|\phi |<20$ and a [26/23] Pad\'e
approximant. For the values of $n$ considered here, 
changing the range of integration has effects smaller
than the errors due to numerical integration (which has an accuracy of 
about 11
significant digits in our calculation) 
provided that we include values up to $|\phi|\simeq 4.9$.
Restricting the range of integration to smaller values produces 
sizable effects. As an example, the small effects due a 
restriction to $|\phi |<4.4$ are shown in Fig. \ref{fig:various}.
Similarly, the values of $a_n$ are not very sensitive to small changes in 
the Pad\'e approximants. Sizable effects are obtained by changing 
the order of the numerator and denominator by approximately 10.
For instance, the effects of using a [14/11] approximant are shown in
Fig. \ref{fig:various}. 

Having demonstrated that we can calculate the first 20 coefficients
$a_n$, at leading order 
in the $1/N$ expansion, 
with at least 10 significant digits, we can now discuss the 
errors associated with other procedures mentioned in the introduction. 
The first procedure consists in truncating $U_0^{\star}$ keeping only the 
terms up to order $(\phi^2)^3$. 
This is procedure inspired by 
perturbation theory amounts to keep only the relevant and 
marginal directions near the Gaussian fixed point. 
From Fig. \ref{fig:various}, we 
see that this procedure generates errors which are of the same order
as the errors due to the use of the leading $1/N$ approximation.
Consequently, this procedure is quite unsuitable to study the 
correction to this approximation. Slightly better results are 
obtained by keeping as many terms as possible in the expansion 
(up to 50 in our calculation) but restricting the range of integration
in such way that we stay within the radius of convergence. Given 
the rescaling of Eq. (\ref{eq:u0}) this means that for $N=5$, we need 
to restrict the integration to $|\phi|< \sqrt{5\times 2.5}\simeq 3.54$
which is substantially smaller than the 
acceptable field cutoff 4.9 mentioned above.
As one can see from Fig. \ref{fig:various}, this creates errors which 
are between one and two orders of magnitude smaller than the $1/N$ 
corrections. This is better but it compares poorly with what can reached 
with Pad\'e approximants.

Again, except for the comparison with independent numerical 
calculations at finite 
$N$, the same calculations can be performed for the SCM with minor changes.
Results very similar to those shown in Fig. \ref{fig:various} for 
the HM can be produced. Since it contains essentially the same 
information, it has not been displayed. It should 
however be noted that 
the number of significant digits obtained with the two alternate procedures 
are lower than in the case of the HM. In the case of the truncation of the 
range of integration, we need to restrict to 
$|\phi|< \sqrt{5\times 0.11}\simeq 0.74$ while a range of about 2 is 
required in order to obtain an accuracy consistent with the method of 
numerical integration.

\section{Conclusions}
We have shown in two differents models where the critical potential
can be calculated at leading order in the $1/N$ expansion that these
potentials have finite radii of convergence due to singularities in 
the complex plane. Do such a results persist at finite $N$?
In the case of the HM, the behavior of the ratios at finite $N$ shown
in Fig. \ref{fig:decayrat} strongly suggests that at large 
real positive $\phi^2$, 
the critical potential still grows like $(\phi^2)^3$. Can an infinite 
sum converging over the entire complex plane have this kind
of behavior? This is certainly not impossible 
(e.g., $(\phi^2)^3+{\rm e}^{-\phi^2}$), however it requires
cancellations that we judge unlikely to happen. Consequently,
we conjecture that the singularities observed are generic 
rather than being an artifact of the large-$N$ limit.

We have observed that in a system of coordinates where 
the HFP can be approximated by polynomials, the procedure 
which consists in considering bare potential truncated at order 
$(\phi^2)^3$ describes the HFP with a 
low accuracy. We are planning to investigate if similar problems 
appear near tricritical fixed points. In  particular, 
reconsidering the RG flows in 
a larger space of bare parameters may affect the generic 
dimension of
the intersections of hypersurface of various codimensions and help us 
finding a more general realization of spontaneous breaking of 
scale invariance with a
dynamical generation of mass.

Our results have qualitative similarities common with those 
of Refs. \cite{pathos}:
we found some ``pathologies'' which force us to look at the 
RG transformations in a more open-minded way.
We are planning \cite{prog} 
to compare in more detail, the leading order results 
presented here with finite $N$ results, as suggested in Ref. 
\cite{comellas} for the local potential approximation.
Another 
issue regarding the $O(N)$ models and which would deserve a more 
detailed investigation 
is the question of first order
phase transitions\cite{hazenfratz,van}.

\begin{acknowledgments}
We thank the Theory group of Fermilab for its hospitality while 
this work was completed and especially B. Bardeen for conversations 
about dynamical mass generation.
This research was supported in part by the Department of Energy
under Contract No. FG02-91ER40664.
\end{acknowledgments}

\end{multicols}

\begin{thebibliography}{10}

\bibitem{wilson}
K. Wilson, Phys.\ Rev.\ D  {\bf 6}, 419 (1972).
\bibitem{ma73}
E. Ma, Rev. Mod. Phys. {\bf 45}, 589 (1973).
\bibitem{col}
S. Coleman, R. Jackiw and H. Politzer, Phys. Rev. D {\bf 10} 2491 (1974).
\bibitem{town}
P. Townsend, Phys. Rev. D {\bf 14} 1715 (1976).
\bibitem{app}
T. Appelquist and U. Heinz, 
Phys. Rev. D {\bf 24} 2169 (1981).
\bibitem{pisarski}
R. Pisarski, Phys. Rev. Lett. {\bf 48}, 574 (1982).
\bibitem{bmb}
W. Bardeen, M. Moshe and M. Bander,  Phys. Rev. Lett. {\bf 52}, 1188 (1984).
\bibitem{amit84}
D. Amit and E. Rabinovici, Nucl. Phys. {\bf B 257}, 371 (1985).
\bibitem{bardeen}
W. Bardeen, C. Leung and S. Love,  Phys. Rev. Lett. {\bf 56}, 1230 (1986);
W. Bardeen and M. Moshe, Phys.\ Rev.\ D  {\bf 34}, 1229 (1986);
W. Bardeen, Fermilab preprint CONF-88/149-T (1988).
\bibitem{david84}
F. David, D. Kessler and H. Neuberger, Phys. Rev. Lett. {\bf 53}, 2071 (1984);
F. David, D. Kessler and H. Neuberger, Nucl. Phys. {\bf B 257}, 695 (1985).
\bibitem{exact}
There is a large literature on this question. Extensive lists of references 
can be found in two recent reviews: 
C. Bagnuls and C. Bervillier, Phys. Reports {\bf 348},
91 (2001); J. Berges, N. Tetradis and C. Wetterich,  Phys. Reports {\bf 362},
223 (2002).
\bibitem{dyson}
F. Dyson, Comm.\ Math.\ Phys.\ {\bf 12}, 91 (1969).
\bibitem{baker}
G. Baker, Phys.\ Rev.\ B{\bf 5}, 2622 (1972). 

\bibitem{loeffel69}
J. Loeffel, A. Martin, B. Simon, and A. Wightman, Phys. Lett. B {\bf 30},  656
  (1969).

\bibitem{kw}
H. Koch and P. Wittwer, Comm. Math. Phys. {\bf 164}, 627 (1994).
\bibitem{guide}
J. Godina, Y. Meurice, and M. Oktay, Phys. Rev. D {\bf 57}, 6326 (1998).
\bibitem{convpert}
Y. Meurice, Phys. Rev. Lett. {\bf 88}, 141601 (2002).

\bibitem{gam3}
J. Godina, Y. Meurice, and M. Oktay, Phys. Rev. D {\bf 57}, R6581 (1998);
J. Godina, Y. Meurice, and M. Oktay, Phys. Rev. D {\bf 59},  096002  (1999).
\bibitem{osc}
Y. Meurice, G. Ordaz and S. Niermann, J. Stat. Phys. {\bf 87}, 363 (1997).
\bibitem{osc1}
Y. Meurice, G. Ordaz and V. G. J. Rodgers, Phys. Rev. Lett. {\bf 75}, 
4555 (1995).
%
\bibitem{baker96}
G. Baker and P. Graves-Morris, {\em Pad\'e Approximants} (Cambridge University
  Press, Cambridge, 1996).
\bibitem{pathos}
A. van Enter, R. Fernandez and A. Sokal, J. Stat. Phys. {\bf 72}, 879 (1993);
A. van Enter and  R. Fernandez, Phys. Rev. E {\bf 59}, 5165 (1999). 
\bibitem{prog}
J. J. Godina, L. Li, Y. Meurice and B. Oktay, work in progress.
\bibitem{comellas}
J. Comellas and A. Travesset, Nucl. Phys. {\bf B 498}, 539 (1997).
\bibitem{hazenfratz}
P. Hasenfratz and A. Hasenfratz, Nucl. Phys. {\bf B 295}, 1 (1988).
\bibitem{van}
A. van Enter and S. Shlosman, cond-mat/0205455.
\end{thebibliography}
\end{document}